\documentclass[12pt]{iopart}
\usepackage{iopams}
\usepackage{setstack}
\usepackage{graphicx}
\usepackage{epsf}

\begin{document}

\newcommand{\eps}{\varepsilon}
\newcommand{\lmax}{l_{\rm max}}

\sloppy

\jl{2}

\title[Positron binding to negative ions]
{Many-body theory calculations of positron binding to negative ions}

\author{J A Ludlow\footnote[1]{Present address: Department of Physics, 
Auburn University, Auburn, AL 36849, USA }
and G F Gribakin}

\address{Department of Applied Mathematics and 
Theoretical Physics, Queen's University Belfast, Belfast BT7 1NN,
Northern Ireland, UK}

\ead{\mailto{ludlow@physics.auburn.edu}, \mailto{g.gribakin@qub.ac.uk}}


\begin{abstract}
A many-body theory approach developed by the authors
[{\it Phys. Rev. A} {\bf 70} 032720 (2004)] is applied to positron bound states
and annihilation rates in atomic systems. Within the formalism, full account
of virtual positronium (Ps) formation is made by summing the electron-positron
ladder diagram series, thus enabling the theory to include all important
many-body correlation effects in the positron problem. Numerical calculations
have been
performed for positron bound states with the hydrogen and halogen negative
ions, also known as Ps hydride and Ps halides. The Ps binding energies of
1.118, 2.718, 2.245, 1.873 and 1.393 eV and annihilation rates of 2.544,
2.482, 1.984, 1.913 and 1.809 ns$^{-1}$, have been obtained for PsH, PsF,
PsCl, PsBr and PsI, respectively.
\end{abstract}

\pacs{36.10.Dr, 71.60.+z, 78.70.Bj, 82.30.Gg}

\maketitle

\section{Introduction}\label{Intr3}

A many-body theory approach developed by the authors (Gribakin and Ludlow
2004) takes into account all main correlation effects in positron-atom
interactions. These are polarization of the atomic system by the positron,
virtual positronium formation and enhancement of the electron-positron
contact density due to their Coulomb attraction. The first two effects
are crucial for an accurate description of positron-atom scattering and
the third one is very important for calculating positron annihilation rates.
In this paper we apply our method to calculate the energies and annihilation
rates for positron bound states with the hydrogen and halogen negative ions.

Positron bound states can have a strong effect on positron annihilation
with matter. For example, positron bound states with molecules give rise
to vibrational Feshbach resonances, which leads to a strong enhancement
of the positron annihilation rates in many polyatomic molecular
gases (Gribakin 2000, 2001, Gilbert \etal 2002, Gribakin \etal 2010).
Nevertheless, the question of positron binding with neutral
atoms or molecules has been answered in the affirmative only recently,
and the ubiquity of such states is only becoming clear now (Dzuba \etal 1995,
Ryzhikh and Mitroy 1997, Mitroy \etal 2001, 2002, Danielson 2009).
On the other hand, it has been known for many decades that positrons bind to
negative ions. Beyond the electron-positron bound
state, or positronium (Ps), the simplest atomic system capable of
binding the positron is the negative hydrogen ion (Ore 1951). The binding
energy, annihilation rate and structure of the resulting compound,
positronium hydride (PsH) have now been calculated to very high precision,
e.g., by variational methods (Frolov \etal 1997). The
information available for heavier negative ions with many valence electrons
is not nearly as accurate (see Schrader and Moxom 2001 for a useful
review). Positronium halides have received most of the attention, with some
calculations dating back fifty years; see, e.g., the Hartree-Fock
calculations of Simons (1953) and Cade and Farazdel (1977), quantum Monte-Carlo 
work by Schrader \etal (1992a, 1993), and more recent configuration
interaction results (Saito 1995, 2005, Saito and Hidao 1998,
Miura and Saito 2003).

At first glance, the physics of positron binding to negative ions is much
simpler than that of positron binding to neutrals. The driving force here is
the Coulomb attraction between the particles. However,
in contrast with atoms, the electron binding energy in a negative ion (i.e.,
the electron affinity, ${\rm EA}$) is always smaller than the binding
energy of Ps, $|E_{1s}|\approx 6.8$~eV. This means that the lowest dissociation
threshold in the positron-anion system is that of the neutral atom and Ps,
so that the positron bound to a negative ion may still escape by Ps emission.
Hence, for the system to be truly bound, its energy should lie {\em below}
the Ps-atom threshold, and the positron energy in the bound state,
$\eps _0$, must satisfy
\begin{equation}\label{EA}
|\eps _0|>|E_{1s}|-{\rm EA} .
\end{equation}
This situation is similar to positron binding to atoms with ionization
potentials smaller than 6.8~eV. The structure of these bound states
is characterized by a large ``Ps cluster'' component
(Mitroy \etal 2002).

The proximity of the Ps threshold in anions means that to yield accurate
binding energies, the method used must account for virtual Ps formation.
The positron also polarizes the electron cloud, inducing an attractive
polarization potential. It has the form $-\alpha e^2/2r^4$ at large
positron-target separations $r$, where $\alpha $ is the dipole polarizability
of the target. The method should thus be capable of describing many-electron
correlation effects. The halogen negative ions have
$np^6$ valence configurations, and are similar to noble-gas atoms. Here
many-body theory may have an advantage over few-body methods (Dzuba \etal 1996).

The many-body theory approach developed by the authors (Gribakin
and Ludlow 2004), employs B-spline basis sets. This enables the sum of the
electron-positron ladder diagram sequence, or vertex function, to be found
exactly. This vertex function accounts for virtual Ps formation. It is
incorporated into the diagrams for the positron correlation potential and
correlation corrections to the
electron-positron annihilation vertex. To ensure convergence with respect
to the maximum orbital angular momentum of the intermediate electron and
positron states in the diagrams, $\lmax $, we use extrapolation.
It is based on the known asymptotic behaviour of the energies and annihilation
rates as functions of $\lmax$ (Gribakin and Ludlow 2002).

In Gribakin and Ludlow (2004) the theory has been successfully applied to
positron scattering and annihilation on hydrogen below the Ps formation
threshold, where the present formalism is exact. This theory is now extended
to treat the more difficult problem of positron interaction with multielectron
atomic negative ions. We test the method for H$^-$, and consider the
halogen negative ions F$^-$, Cl$^-$, Br$^-$ and I$^-$. Note that a conventional
notation for the positron bound with a negative ion $A^-$ is Ps$A$, rather
than $e^+A^-$, hence one has PsH, PsF, etc. (Schrader 1998).

\section{Calculation of positron binding using Dyson's equation}
\label{sec:dyson}

The many-body theory method for positron bound states is similar to that
developed for electron-atom binding in negative ions (Chernysheva \etal 1988,
Johnson \etal 1989, Gribakin \etal 1990, Dzuba \etal 1994), and used for
positron-atom bound states by Dzuba \etal (1995).

The Green function of the positron interacting with a many-electron
system (``target'') satisfies the Dyson equation (see, e.g., Migdal 1967),
\begin{equation}\label{eq:Dys}
(E-\hat H_0)G_E({\bf r},{\bf r}')-\int \Sigma _E({\bf r},{\bf r}'')
G_E({\bf r}'',{\bf r}')\rmd {\bf r}''=\delta ({\bf r}-{\bf r}'),
\end{equation}
where $\hat H_0$ is the zeroth-order positron Hamiltonian, and $\Sigma _E$ is
the {\em self-energy}. A convenient choice of $\hat H_0$ is that of the
positron moving in the field of the Hartree-Fock (HF) target ground state.
The self-energy then
describes the correlation interaction between the positron and the target beyond
the static-field HF approximation (Bell and Squires 1959). It can
be calculated by means of the many-body diagrammatic expansion in powers of
the electron-positron and electron-electron Coulomb interactions (see below).

If the positron is capable of binding to the system, i.e., the target has a
positive positron affinity ${\rm PA}$, the positron Green function
$G_E({\bf r},{\bf r}')$ has a pole at $E=\eps_0\equiv -{\rm PA}$,
\begin{equation}\label{eq:pole}
G_E({\bf r},{\bf r}')\underset{E\rightarrow\eps_0}{\simeq }
\frac{\psi_0({\bf r})\psi _0^*({\bf r}')}{E-\eps_0} .
\end{equation}
Here $\psi _0({\bf r})$ is the quasiparticle wavefunction which describes
the bound-state positron. It is equal to the projection of the total 
ground-state wavefunction of the positron and $N$ electrons,
$\Psi _0({\bf r}_1,\dots,{\bf r}_N,{\bf r})$,
onto the target ground-state wavefunction, 
$\Phi _0({\bf r}_1,\dots,{\bf r}_N)$ ,
\begin{equation}\label{eq:quasi}
\psi_0({\bf r})=\int \Phi_0^*({\bf r}_1,\dots,{\bf r}_N)
\Psi_0({\bf r}_1,\dots,{\bf r}_N,{\bf r})
\rmd {\bf r}_1\dots \rmd {\bf r}_N .
\end{equation}
The normalization integral for $\psi_0({\bf r})$,
\begin{equation}\label{eq:norm}
a=\int|\psi_0({\bf r})|^2\rmd {\bf r}<1 ,
\end{equation}
can be interpreted as the probability that the electronic subsystem of the
positron-target complex remains in its ground state.

The magnitude of $a$ quantifies the extent to which the structure of the
complex is that of the positron bound to the anion, $e^+A^-$, as opposed to
that of the Ps atom orbiting the neutral atom. Such separation is a feature
of the ``heuristic wavefunction model'' (Mitroy \etal 2002). If the former
component dominates the wavefunction, the value of $a$ is expected to be be
close to unity. If, on the other hand, the wavefunction contains a large
``Ps cluster'' component, the value of $a$ will be notably smaller.

By taking the limit $E\rightarrow \eps _0$ in equation (\ref{eq:Dys}),
one obtains the Dyson equation for the quasiparticle wavefunction
$\psi_0({\bf r})$ (``Dyson orbital'') and the bound-state energy $\eps_0$,
\begin{equation}\label{eq:Dyson}
\hat H _0\psi_0({\bf r}) + \int\Sigma _{\eps_0}({\bf r},{\bf r}')
\psi_0({\bf r}')\rmd {\bf r}'=\eps _0\psi_0({\bf r}) .
\end{equation}
This equation is analogous to the standard Schr\"odinger eigenvalue problem,
except that the correlation potential $\Sigma $ depends on the energy.

The eigenstates $\varphi _\eps ({\bf r})$ of the Hamiltonian $\hat H_0$,
\begin{equation}\label{eq:HF}
\hat H _0\varphi _\eps ({\bf r})=\eps \varphi _\eps ({\bf r}) ,
\end{equation}
are characterized by their energies $\eps $ and the orbital angular
momentum $l$, implicit in this notation. They form a complete single-particle
positron basis set. For negative ions the spectrum of $\hat H_0$ consists of
both discrete (Rydberg) and continuum states. This set can be used to expand
the quasiparticle bound-state wavefunction,
\begin{equation}\label{eq:expand}
\psi _0({\bf r})=\sum _\eps C_\eps \varphi _\eps ({\bf r}),
\end{equation}
and cast the Dyson equation (\ref{eq:Dyson}) in the matrix form,
\begin{equation}\label{eq:Dyson1}
\eps C_{\eps }+\sum_{\eps '}
\langle \eps |\Sigma_{\eps_0}|\eps '\rangle C_{\eps '}
=\eps_0C_{\eps },
\end{equation}  
where the sums in equations (\ref{eq:expand}) and (\ref{eq:Dyson1}) include the
positive-energy continuum, as well as the discrete negative-energy states,
and
\begin{equation}\label{eq:Sigmat}
\langle \eps|\Sigma _E|\eps '\rangle =\int \varphi_ \eps ^*({\bf r})
\Sigma _E({\bf r},{\bf r}')\varphi _{\eps '}({\bf r}')\rmd {\bf r}
\rmd {\bf r}'.
\end{equation} 

In practice the continuum is discretized by using a B-spline basis set (see
below), and equation (\ref{eq:Dyson1}) becomes a matrix eigenvalue
problem. Therefore, to find $\eps_0$ and $C_{\eps }$ one simply
needs to diagonalize the matrix,
\begin{equation}\label{Dyson4}
\eps_{\eps }\delta_{\eps \eps '}+
\langle\eps |\Sigma _E|\eps '\rangle .
\end{equation}
Its lowest eigenvalue $\eps_0(E)$ depends on the energy $E$ at which
$\Sigma_E$ is calculated, and the diagonalization must be repeated several
times until self-consistency is achieved: $\eps_0(E)=E$. Knowing the
dependence of the eigenvalue on $E$ allows one to determine the normalization
integral, equation (\ref{eq:norm}), via the relation (Migdal 1967),
\begin{equation}\label{norm1}
a=\left(1-\left.\frac{\partial\eps_0(E)}{\partial E}
\right|_{E=\eps_0}\right)^{-1} .
\end{equation}
Note that owing to the spherical symmetry of the target, the states
$\varphi _\nu $ and $\psi _0$ have definite orbital angular momenta. To find
the bound positron ground state it is sufficient to calculate the self-energy
matrix (\ref{eq:Sigmat}) for the $s$-wave positron only.

\begin{figure}[ht]
\begin{center}
\includegraphics*[width=8cm]{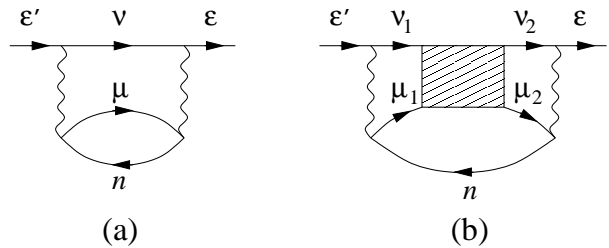}
\end{center}
\caption{Main contributions to the positron self-energy. Diagram (a) describes
the effect of polarization in the lowest, second order, $\Sigma ^{(2)}$; diagram
(b) accounts for virtual Ps formation, $\Sigma ^{(\Gamma )}$. Top lines in
the diagrams describe the positron. Other lines with the arrows to the right
are excited electron states, and to the left -- holes, i.e., electron states
occupied in the target ground state. Wavy lines are the Coulomb interactions.
Summation over the intermediate  positron, electron and hole states is carried
out.}
\label{fig:Sig_2Ps}
\end{figure}

\begin{figure}[ht]
\begin{center}
\includegraphics*[width=15.7cm]{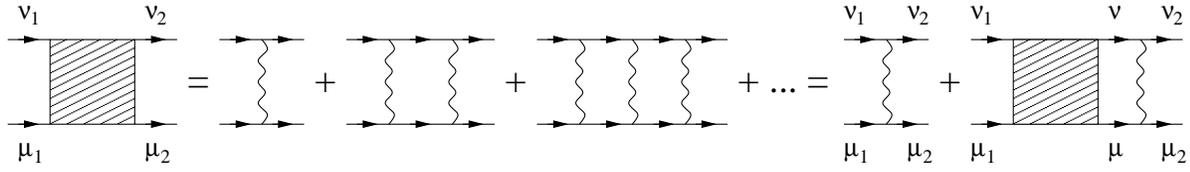}
\end{center}
\caption{Electron-positron ladder diagram series and its sum, the vertex
function $\Gamma $ (shaded block). Comparison between the left- and right-hand
sides of the diagrammatic equation shows that $\Gamma $ can be found by
solving a linear matrix equation.}
\label{fig:lad}
\end{figure}

The accuracy of the binding energy obtained from the Dyson equation
depends upon the accuracy to which the self-energy has been determined.
As mentioned in the Introduction, polarization of the target and virtual Ps
formation are the two most important effects that need to be accounted for.
The effect of target polarization is described in the
leading order by the 2nd-order diagram $\Sigma ^{(2)}$,
figure \ref{fig:Sig_2Ps}(a). Following Gribakin and Ludlow (2004), the Ps
formation contribution $\Sigma ^{(\Gamma )}$ shown in
figure~\ref{fig:Sig_2Ps}(b), is obtained by summing the electron-positron
ladder diagram series to all orders, figure~\ref{fig:lad}. This procedure
amounts to calculation of the electron-positron vertex function $\Gamma $
shown by the shaded block. Analytical expressions for the
diagrams can be found in Gribakin and Ludlow (2004).

\begin{figure}[ht]
\begin{center}
\includegraphics*[width=15.7cm]{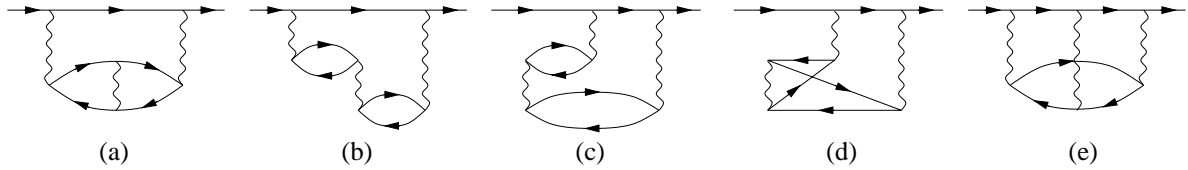}
\end{center}
\caption{Third-order contributions to the positron self-energy,
$\Sigma ^{(3)}$. Mirror images of the diagrams (c) and (d) are also 
included. The top line describes the positron.}
\label{fig:Sig_3}
\end{figure}

For hydrogen, the self-energy is given exactly by the two diagrams
in figure \ref{fig:Sig_2Ps}, $\Sigma = \Sigma ^{(2)} + \Sigma ^{(\Gamma )}$,
provided the intermediate electron and positron states are calculated in the
field of the bare nucleus (Gribakin and Ludlow 2004). For multi-electron
targets one may also consider higher-order corrections not included in the
virtual Ps contribution, figure \ref{fig:Sig_2Ps}(b). In the present
calculation a set of 3rd-order diagrams shown in figure \ref{fig:Sig_3} will
be included.
Diagrams (a), (b), (c) and (d) represent corrections to the 2nd-order
polarization diagram, of the type described by the random-phase
approximation with exchange (RPAE, Amusia \etal 1975). They account for the
electron-hole interaction and screening of the positron Coulomb field, and
correct the value of the dipole polarizability $\alpha $ of the target.
Diagram (e) accounts for the positron-hole
repulsion. The contribution of the diagrams in figure \ref{fig:Sig_3} is
denoted collectively by $\Sigma ^{(3)}$. Calculation of these diagrams
will allow us to gauge the importance of these corrections, and
even to include effectively higher-order diagrams (see below).

\begin{figure}[ht]
\begin{center}
\includegraphics*[width=5cm]{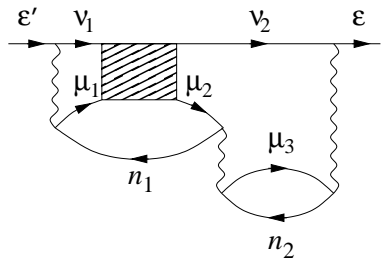}
\end{center}
\caption{Screening correction to the virtual Ps contribution
$\Sigma ^{(\Gamma )}$.}
\label{fig:sig_Gam_cor}
\end{figure}

Of course, any many-body theory calculation can at best include only dominant
classes of diagrams, leaving out an infinite number of other higher-order
diagrams. For example, the diagram in figure \ref{fig:sig_Gam_cor}
has the effect of screening the positron-electron interaction accompanying
virtual Ps formation in $\Sigma^{(\Gamma)}$, figure \ref{fig:Sig_2Ps}(b).

The number of diagrams increases rapidly as one moves to higher orders. The
effort required to evaluate these diagrams becomes prohibitive relative to
their small contribution. It would therefore be useful to find a simple
method to estimate the contribution of the higher-order diagrams.
This will allow us to take into account the effect of electron screening
beyond the corrections shown in figure \ref{fig:Sig_3}.

A useful quantity for estimating the size of a contribution to the self-energy
$\Sigma $ is a dimensionless measure of its ``strength'' (Dzuba \etal 1994), 
\begin{eqnarray}\label{strength}
g_E(\Sigma)&=&\int G^{(0)}({\bf r}',{\bf r})\Sigma_E({\bf r},{\bf r}')
{\rm d}{\bf r}{\rm d}{\bf r}'\\
&=&-\sum_{\nu}\frac{\langle \eps _\nu|\Sigma_E|\eps _\nu\rangle}
{\eps_{\nu}} , 
\end{eqnarray}  
where $G^{(0)}$ is the 0th order positron Green function calculated at $E=0$. 
Let
\begin{equation}\label{Str}
S_E=\frac{g_E(\Sigma^{(3)})}{g_E(\Sigma^{(2)})}
\end{equation}
be the ratio of the strength of the sum of the 3rd-order polarization diagrams
(figure \ref{fig:Sig_3}) to the strength of the 2nd-order polarization diagram. 
This quantity $S_E$ can then be used to estimate higher-order contributions
to $\Sigma^{(\Gamma)}$.

As a check, we test that the binding energy obtained using
$\Sigma^{(2+3)}=\Sigma^{(2)}+\Sigma^{(3)}$ is close to that obtained with
$\Sigma^{(2)}$ multiplied by $1+S_E$, i.e., using
\begin{equation}\label{eq:23prime}
(1+S_E)\Sigma^{(2)}\equiv \Sigma^{(2+3')}.
\end{equation}
An estimate of the {\em total} self-energy corrected for the screening effects
in the lowest order can then be obtained as
\begin{equation}\label{eq:sig_cor}
(1+S_E)\left[ \Sigma^{(2)}+\Sigma ^{(\Gamma )}\right] .
\end{equation}

\begin{figure}[ht]
\begin{center}
\includegraphics*[width=6.5cm]{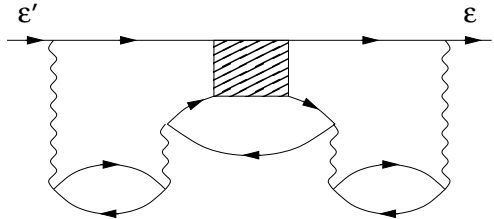}
\end{center}
\caption{Higher-order screening correction to the virtual Ps contribution
$\Sigma ^{(\Gamma )}$, cf. figure~\ref{fig:sig_Gam_cor}.}
\label{fig:sig_Gam_cor2}
\end{figure}

As we will see in section \ref{sec:num}, the relative effect of screening is
negative, $S_E<0$, which means that screening reduces the magnitude of the
self-energy. Similar higher-order terms in the self-energy expansion will
alternate in sign. Thus, for example, a diagram such as that shown in
figure~\ref{fig:sig_Gam_cor2}, will tend to compensate the lowest-order
screening correction in figure \ref{fig:sig_Gam_cor}.

Assuming that the sequence of screening corrections behaves like a geometric
series, its effect can approximately be taken into account by using the
{\em screened} self-energy, which we denote $\Sigma^{\rm (scr)}$:
\begin{equation}\label{eq:Sig_scr}
\Sigma^{\rm (scr)}=\frac{1}{1-S_E}\left[ \Sigma^{(2)}+\Sigma ^{(\Gamma )}\right].
\end{equation}
We expect that this approximation should yield our best prediction for the
binding energy.

\section{Calculation of the positron annihilation rate in the bound state}
\label{sec:ann}

The spin-averaged positron annihilation rate $\Gamma_a$ in the bound state
can be expressed in terms of the average contact electron-positron density
$\rho_{ep}$ (see, e.g., Berestetskii \etal 1982)
\begin{equation}\label{Zeff3}
{\Gamma_a}={\pi}r_0^2c\rho_{ep} ,
\end{equation}
where $r_0$ is the classical radius of the electron, $c$ is the speed of
light, and $\rho_{ep}$ is given by the integral,
\begin{equation}\label{Zeff31}
\rho_{ep}=\sum_{i=1}^N\int
\left|\Psi _0({\bf r}_1,{\bf r}_2,\dots,{\bf r}_N,{\bf r})\right|^2
\delta({\bf r}-{\bf r}_i){\rm d}{\bf r}_1
\dots{\rm d}{\bf r}_N{\rm d}{\bf r} ,
\end{equation} 
where $\Psi_0({\bf r}_1,{\bf r}_2,\dots,{\bf r}_N,{\bf r})$ is the full
$(N+1)$-particle bound-state wavefunction of the $N$ electron coordinates
${\bf r}_i$ and positron coordinate ${\bf r}$.

\begin{figure}[ht]
\begin{center}
\includegraphics*[width=15cm]{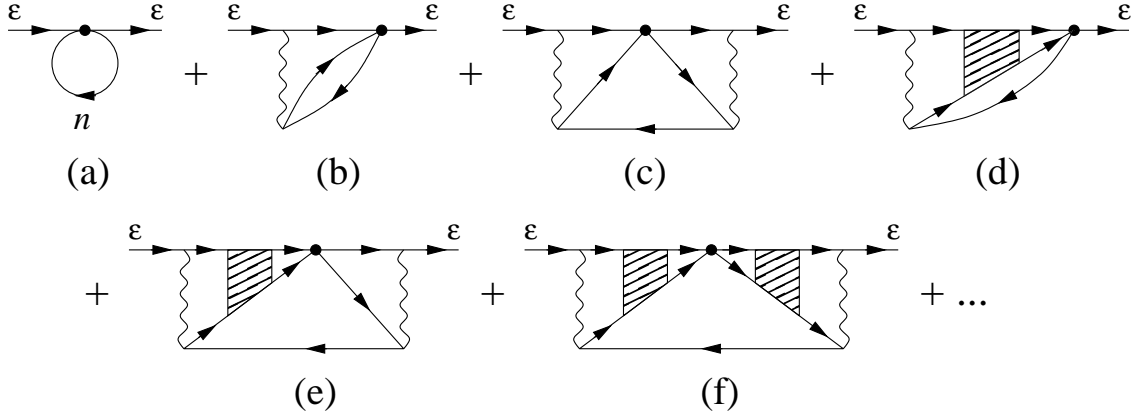}
\end{center}
\caption{Many-body theory expansion for the contact electron-positron density.
The solid circle in the diagrams is the delta-function annihilation vertex,
cf. equation (\ref{Zeff31}). Diagrams (b), (d) and (e) are
multiplied by two to account for their mirror images.}
\label{fig:diagHz}
\end{figure}

Figure \ref{fig:diagHz} shows a series of diagrams that would 
constitute a complete set of annihilation diagrams for a one-electron system
(Gribakin and Ludlow 2004). Here $\eps $ represents the positron bound-state
Dyson orbital $\psi _0({\bf r})$ normalized as per equation~(\ref{eq:norm}).   

Diagram (a) in figure~\ref{fig:diagHz} is the overlap of the positron
and HF electron densities. It represents the independent-particle
approximation to the annihilation vertex with the
contact density
\begin{equation}\label{eq:rho_0}
\rho_{ep}^{(0)}=\sum _{n=1}^{N}\int |\phi _n({\bf r})|^2|\psi _0({\bf r})|^2
{\rm d}{\bf r},
\end{equation}
where $\phi _n({\bf r})$ is the HF orbital of hole $n$. The first-order
correction, diagram (b), can be thought of as the analogue of $\Sigma^{(2)}$,
and will be denoted by $\rho_{ep}^{(1)}$. The diagrams with the vertex
function $\Gamma $, e.g., (d) in figure ~\ref{fig:diagHz}, are
particularly important in the calculation of $\rho_{ep}$, as the annihilation
takes place at a point, and is strongly enhanced by the Coulomb attraction
in the annihilating electron-positron pair.

\begin{figure}[ht]
\begin{center}
\includegraphics*[width=15cm]{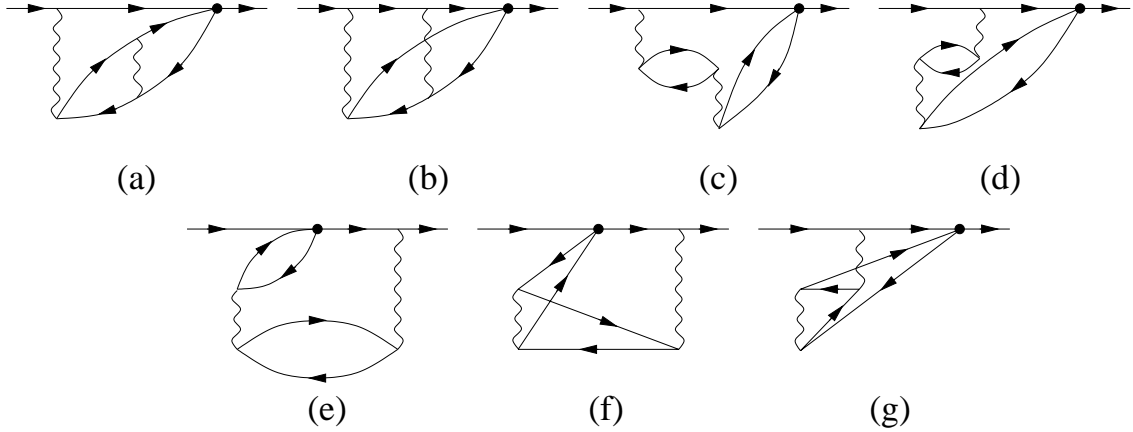}
\end{center}
\caption{Annihilation diagrams with two Coulomb interactions, including those
of RPA-type, $\rho_{ep}^{(2)}$. The top line describes the positron. All the
diagrams have equal mirror images.}
\label{fig:diagz}
\end{figure}

The diagrams shown in figure \ref{fig:diagHz} represent a basic set of
contributions with a single hole line, which one needs to consider to obtain
a reliable answer. We will denote the corresponding density as
$\rho_{ep}^{(0)}+\rho_{ep}^{(\Delta)}$, $\rho_{ep}^{(\Delta)}$ representing a
correction to the zeroth-order contact density. Similar to the self-energy
diagrams in figure~\ref{fig:Sig_2Ps}, they represent the exact answer
for the positron-hydrogen system, provided the electron and positron
intermediate states are calculated in the field of the bare nucleus
(Gribakin and Ludlow 2004). For complex many-electron systems it may be
necessary to account for the effects of electron screening when calculating
$\rho_{ep}$. A series of RPA-type annihilation diagrams is therefore also
calculated, see figure \ref{fig:diagz}.

\begin{figure}[ht]
\begin{center}
\includegraphics*[width=5cm]{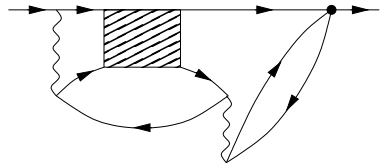}
\end{center}
\caption{Screening correction to the annihilation diagram containing the
vertex function $\Gamma $.}
\label{fig:ann_Gam_cor}
\end{figure}

Similarly, screening corrections to the annihilation diagrams containing
the $\Gamma$ block can also be considered, as shown in
figure~\ref{fig:ann_Gam_cor}.

The diagrams shown in figure \ref{fig:diagz}, $\rho_{ep}^{(2)}$, can be thought
of as next-order corrections to $\rho_{ep}^{(1)}$, diagram (b) in
figure~\ref{fig:diagHz}. By evaluating the ratio 
\begin{equation}\label{Str1}
C=\rho_{ep}^{(2)}/\rho_{ep}^{(1)},
\end{equation}
an estimate can be made of the total contact density $\rho_{ep}$ that includes
higher-order corrections in a manner similar to the self-energy
[cf. equation~(\ref{eq:Sig_scr})],
\begin{equation}\label{Str2}
\rho_{ep}=\rho_{ep}^{(0)}+\frac{1}{1-C}\rho_{ep}^{(\Delta)} .
\end{equation}

\section{Numerical Implementation}\label{sec:num}

The Hartree-Fock ground state of the negative ions is first found. 
The frozen-core HF Hamiltonian for an electron or a positron
(with and without exchange, respectively) is then diagonalized in a
B-spline basis (Sapirstein \etal 1996). The corresponding eigenvectors
provide bases of single-particle electron and positron states, cf.
equation~(\ref{eq:HF}). The spectrum of these states for the electron
includes the negative-energy ground-state orbitals (hole states) and
positive-energy excited states spanning the electron continuum,
see, e.g., figure 6 in Gribakin and Ludlow (2004). The positron basis
contains a number of negative-energy Rydberg states augmented by
the discretized positive energy positron ``continuum''.

The effective spanning of the continuum is achieved by using an exponential
radial knot sequence for the B-splines. For H$^-$, the first 23 eigenstates
generated from a set of 60 splines of order 9 were used, with a box size
of $R=30$~au. For the other systems, namely F$^-$, Cl$^-$, Br$^-$ and I$^-$,
the first 20 states from a set of 40 splines of order 6 were used with
$R=30$~au. Only
the outermost $s$ and $p$ subshells were included when calculating the
self-energy and annihilation diagrams. More strongly bound inner-shell
electrons are only weakly perturbed by the positron. Their contribution
to the correlation potential and annihilation vertex is relatively small,
and has been neglected.

The diagrammatic contributions to the self-energy and contact density
described in sections \ref{sec:dyson} and \ref{sec:ann}, are calculated by
direct summation over the intermediate electron and positron states, and the
vertex function is found by solving a linear matrix equation.
The use of B-spline bases ensures quick convergence with respect to the number
of states with a particular angular momentum $l$ included in the calculation.
In addition, the convergence with respect to the maximum orbital angular
momentum included in the calculation, $\lmax$, also needs to be
considered. This is done by extrapolation through the use of the asymptotic
formulae (Gribakin and Ludlow 2002),
\begin{equation}\label{asym3}
\eps_0=\eps_0^{(\lmax)}-\frac{A}{(\lmax+1/2)^3} ,
\end{equation}
and
\begin{equation}\label{asym4}
\rho_{ep}=\rho_{ep}^{(\lmax)}+\frac{B}{(\lmax+1/2)} ,
\end{equation}
where $\eps_0^{(\lmax)}$ and $\rho_{ep}^{(\lmax)}$ are the bound-state
energy and contact density obtained in a calculation for a given
$\lmax$, and $A$ and $B$ are constants. While the derivation
of equations (\ref{asym3}) and (\ref{asym4}) is based on perturbation theory
(Gribakin and Ludlow 2002), this asymptotic behaviour is confirmed by
nonperturbative many-body-theory calculations (Gribakin and Ludlow 2004), and
by configuration-interaction calculations of positron binding and
annihilation in atoms (Mitroy and Bromley 2006). The constants $A$ and $B$
are found numerically and have different values for each system studied. 

\section{Results}

\subsection{Details of calculations for {\rm PsCl}}
In this section a detailed examination of the results for PsCl will be
presented. This should illustrate how the final results for the other systems
were arrived at.

\begin{figure}[pht]
\begin{center}
\includegraphics*[width=9.0cm]{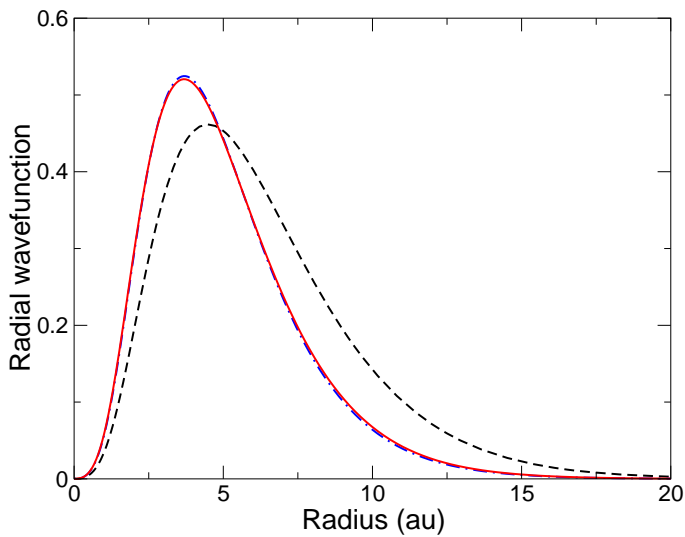}
\end{center}
\caption{Radial positron wavefunction in PsCl: $\dashed $, Hartree-Fock;
$\chain $, Dyson orbital calculated using the self-energy
$\Sigma^{(2+\Gamma+3)}$; $\full $, Dyson orbital obtained with
$\Sigma^{\rm (scr)}$. The HF and Dyson orbitals on the graph are normalized
to unity.}
\label{fig:PsCl}
\end{figure}

The positron radial wavefunctions for PsCl obtained by solving the Dyson
equation using the {\it ab initio} self-energy $\Sigma^{(2+\Gamma+3)}$ and the
screened self-energy $\Sigma^{\rm (scr)}$ are compared to the HF positron
wavefunction in figure \ref{fig:PsCl}. The inclusion of the attractive
correlation potential results in lower energies of the positron bound states,
and hence, more compact wavefunctions. Otherwise the two Dyson orbitals are
very similar to each other.

\begin{figure}[pht]
\begin{center}
\includegraphics*[width=9.0cm]{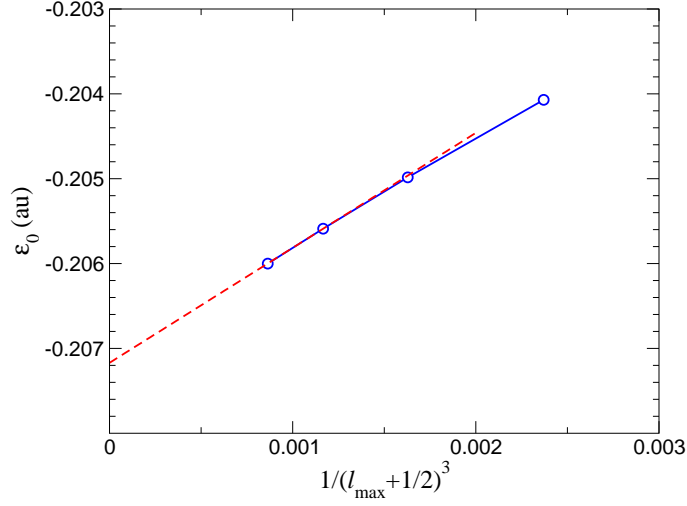}
\end{center}
\caption{Convergence of the binding energy for PsCl as a function of $\lmax $.
Open circles connected by a solid line to guide the eye, show the
energies $\eps _0$ calculated using $\Sigma^{(2+\Gamma+3)}$ at $E=-0.207$ au
for $\lmax =7$--10; dashed line shows extrapolation.}
\label{fig:Clcone}
\end{figure}

When solving the Dyson equation, the self-energy and the bound-state energy
$\eps _0$ were calculated for a number of maximum orbital angular momenta,
e.g., $\lmax=7$--10, and then the asymptotic behaviour (\ref{asym3}) was used
to find the result for $\lmax \rightarrow \infty $. This procedure
is illustrated for PsCl in figure \ref{fig:Clcone}. Extrapolation
from $\lmax=10$ to infinity increases the binding energy by about 0.5\%.

\begin{figure}[pht]
\begin{center}
\includegraphics[width=9.0cm]{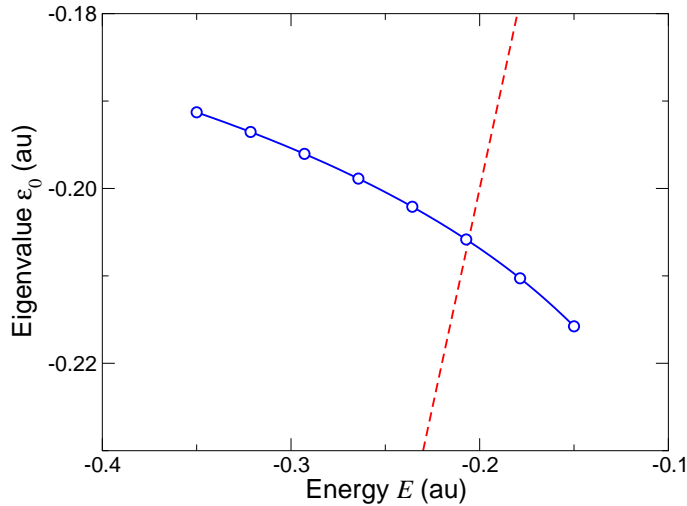}
\end{center}
\caption{Open circles connected by the solid line show the positron energy
$\eps_0(E)$ for PsCl obtained from the Dyson equation with
$\Sigma^{(2+\Gamma+3)}$, as a function of the
energy at which the self-energy was calculated; dashed line is $\eps_0=E$.
The intersection of the two lines, $\eps_0(E)=E$, gives the binding energy.
The gradient of $\eps_0(E)$ at this point is used to calculate the
normalization constant $a$ from equation~(\ref{eq:norm}).}
\label{fig:grad}
\end{figure}

Before the contact density $\rho_{ep}$ can be determined, the positron Dyson
orbital must be correctly normalized via equation~(\ref{norm1}). This is
achieved by calculating the self-energy for a number of energies $E$ and
finding the lowest eigenvalue of the matrix (\ref{Dyson4}) at these energies,
giving $\eps _0(E)$. This is repeated to self-consistency,
$E=\eps _0(E)$, and the gradient $\partial\eps_0(E)/\partial E$ is found
at this point. This is illustrated for PsCl in figure \ref{fig:grad}.

\begin{figure}[pht]
\begin{center}
\includegraphics*[width=9.0cm]{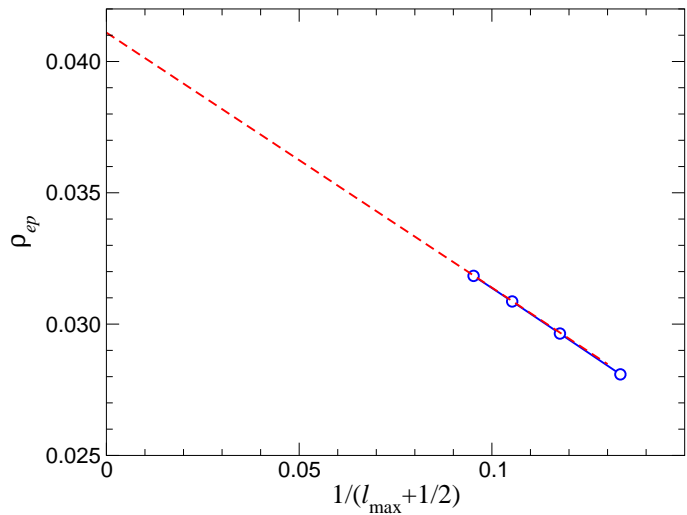}
\end{center}
\caption{Convergence of the electron-positron contact density $\rho_{ep}$ for
PsCl as a function of $\lmax$. Open circles connected by a solid line to guide
the eye, show values of $\rho_{ep}^{(0)}+\rho_{ep}^{(\Delta)}+\rho_{ep}^{(2)}$
obtained for $\lmax =7$--10 in the calculation using $\Sigma^{(2+\Gamma+3)}$;
dashed line shows extrapolation.}
\label{fig:Clconz}
\end{figure}

According to equation~(\ref{asym4}), the electron-positron contact density
$\rho_{ep}$ converges much slower than the energy, and extrapolation with
respect to $\lmax $ is much more important here. This is illustrated
for PsCl in figure~\ref{fig:Clconz}. Extrapolation beyond $\lmax =10$
increases the contact density, and hence, the annihilation rate, by
about 30\%.

The calculations are performed for a number of approximations to the
self-energy. This enables us to determine the relative magnitude of various
diagrams, and helps to clarify which physical effects are important to include
so as to obtain an accurate binding energy. The positron energies obtained
for PsCl  using different approximations are given in table~\ref{table31}.
 
\begin{table}[ht]
\caption{Positron bound-state energies $\eps _0$ (in au) for Cl$^-$
obtained using various approximations to the correlation potential.}
\label{table31}
\begin{indented}
\item[]
\begin{tabular}{@{}ccccccc}
\br
HF & $\Sigma^{(2)}$ & $\Sigma^{(2+\Gamma)}$ & $\Sigma^{(2+3)}$ & 
$\Sigma^{(2+\Gamma+3)}$ &
$\Sigma^{(2+3')}$ & $\Sigma^{\rm (scr)}$ \\
\mr
$-0.1419$ & $-0.1855$& $-0.2276$  & $-0.1663$ & $-0.2072$ &
$-0.1641$ & $-0.1998$ \\
\br
\end{tabular}
\end{indented}
\end{table}

In the HF approximation, the energy of the lowest $s$ wave positron state 
is $-0.1419$~au. The Ps binding energy (BE) is determined from the positron
affinity ${\rm PA}=|\eps_0|$ using equation (\ref{conv}), which gives
${\rm BE} = 0.672$ eV. This value is in agreement with the
HF results of Cade \etal (1977). Therefore, PsCl is bound even in the static HF
approximation. However, the self-energy is essential in determining an
accurate binding energy. The second-order polarization diagram $\Sigma^{(2)}$
increases the binding energy, and the inclusion of the virtual Ps formation
contribution $\Sigma^{(\Gamma)}$ increases it even further. When we add the
3rd-order corrections $\Sigma^{(3)}$ to $\Sigma^{(2)}$, the binding reduces
noticeably. This means that screening of the Coulomb interaction is important.
In particular, the positron binding energy of $0.2276$~au, obtained with
$\Sigma^{(2+\Gamma)}$, becomes equal to $0.2072$~au when the total self-energy
$\Sigma^{(2+\Gamma+3)}$ is used.

However, this calculation neglects the effect of screening on the virtual
Ps-formation contribution $\Sigma^{(\Gamma)}$. Evaluating the magnitude of
screening via equation (\ref{Str}) we obtain $S_E=-0.45$. When the effect of
screening on $\Sigma^{(2)}$ is included via the factor $1+S_E$, as per
equation~(\ref{eq:23prime}), the corresponding result (labelled
$\Sigma^{(2+3')}$) is very close to that obtained with $\Sigma^{(2+3)}$
(see table \ref{table31}). The application of approximation (\ref{eq:Sig_scr}),
denoted $\Sigma^{\rm (scr)}$, gives our best estimate of the positron binding
energy, 0.1998~au, corresponding to a Ps binding energy of 2.245~eV. This
value is only slightly below the completely {\it ab initio} value of 2.437~eV
obtained using $\Sigma^{(2+\Gamma+3)}$.

The contact densities calculated using the Dyson orbitals obtained
with $\Sigma^{(2+\Gamma+3)}$ and $\Sigma^{\rm (scr)}$ are quite close, as are the
energies and wavefunctions.  We show the breakdown of the contributions to
$\rho_{ep}$ in table \ref{table32}. To appreciate the scale of densities
involved, it is useful to remember that the contact density of ground state Ps
is $\rho_{ep}({\rm Ps})=1/8\pi\approx 0.0398$~au. Note that although
$\Sigma^{(2+\Gamma+3)}$ gives a slightly larger binding energy and a more
compact positron wavefunction than $\Sigma^{\rm (scr)}$, the densities obtained
in the former approximation are lower. This is due to a smaller normalization
constant $a$, which results from a somewhat stronger energy dependence of
$\Sigma^{(2+\Gamma+3)}$.
  
\begin{table}[ht]
\caption{Breakdown of contributions to the electron-positron contact density
in PsCl (in au).}
\label{table32} 
\begin{indented}
\item[]
\begin{tabular}{@{}ccccccc}
\br
Approx. & $\rho_{ep}^{(0)}$ & $\rho_{ep}^{(1)}$ &
$\rho_{ep}^{(0)}+\rho_{ep}^{(\Delta)}$  & $\rho_{ep}^{(2)}$ & total &
$\rho _{ep}$, eq. (\ref{Str2}) \\
\mr
$\Sigma^{(2+\Gamma+3)}$ & 0.00841 & 0.00931 & 0.04263 & $-0.00155$ &
0.04108 & $-$ \\
$\Sigma^{\rm (scr)}$  & 0.00872 & 0.00964 & 0.04444 & $-0.00162$ &
0.04281 & 0.03929 \\
\br
\end{tabular}
\end{indented}
\end{table}

The zeroth-order diagram, $\rho_{ep}^{(0)}$, gives only about 20\% of the total
density, with $\rho_{ep}^{(1)}$ giving another 20\% and the rest coming from
higher order diagrams in $\rho_{ep}^{(\Delta)}$ (figure \ref{fig:diagHz}). As
with the self-energy, the inclusion of screening effects ($\rho_{ep}^{(2)}$,
figure \ref{fig:diagz}) reduces the total. However, the effect of screening 
on the annihilation vertex is much smaller than that on the correlation
potential, as indicated by the value of $C=-0.17$, equation (\ref{Str1}).
Physically, this is related to the fact that in the annihilation vertex
corrections, small electron-positron separations dominate. 
Finally, using equation (\ref{Str2}) to account for the effect of screening on
the diagrams in $\rho_{ep}^{(\Delta)}$, we obtain our best prediction
for the contact density (last column in table~\ref{table32}). This corresponds
to the PsCl decay rate of 1.984~ns$^{-1}$, which is close to the spin-averaged
decay rate of Ps, 2.01 ns$^{-1}$.

\subsection{Results for {\rm PsH}, {\rm PsF}, {\rm PsCl}, {\rm PsBr} and
{\rm PsI}}

The final results for PsH, PsF, PsCl, PsBr and PsI obtained with the
correlation potential $\Sigma^{(2+\Gamma+3)}$ and density
$\rho_{ep}^{(0)}+\rho_{ep}^{(\Delta)}+\rho_{ep}^{(2)}$, and $\Sigma^{\rm (scr)}$
and screened densities from equation~(\ref{Str2}), are shown in
table \ref{table2}. In all cases the positron is bound in the $s$-wave, all
higher lying quasi-bound states being unstable against Ps emission.
Note that the latter is true for the electron-spin-singlet states, as
excited ``unnatural parity'' electron-spin-triplet Ps-atom
bound states have been discovered recently for the hydrogen and the alkalis
(Mitroy and Bromley 2007, Mitroy \etal 2007).

\begin{table}[ht]
\caption{Positron binding energies, normalization constants and
contact densities for PsH and Positronium halides.}
\label{table2}
\begin{indented}
\item[]
\begin{tabular}{@{}lcccccc}
\br
Compound & $\eps_0$$^{\rm a}$ au & $a$$^{\rm a}$ & $\rho_{ep}$$^{\rm a}$ & 
$\eps_0$$^{\rm b}$ au & $a$$^{\rm b}$ & $\rho_{ep}$$^{\rm b}$ \\
\mr
PsH & $-0.27619$ & 0.714 & 0.05231 & $-0.26338$ & 0.748 & 0.05037 \\
PsF & $-0.22778$ & 0.950 & 0.04790 & $-0.22489$ & 0.958 & 0.04913 \\
PsCl & $-0.20718$ & 0.875 & 0.04108 & $-0.19975$ & 0.894 & 0.03929 \\
PsBr & $-0.20373$ & 0.834 & 0.03910 & $-0.19523$ & 0.868 & 0.03788 \\
PsI & $-0.19805$ & 0.794 & 0.03707 & $-0.18878$ & 0.835 & 0.03582 \\
\br
\end{tabular}
\item[$^{\rm a}$]~Dyson equation solved using $\Sigma^{(2+\Gamma+3)}$,
density $\rho _{ep}=\rho_{ep}^{(0)}+\rho_{ep}^{(\Delta)}+\rho_{ep}^{(2)}$.
\item[$^{\rm b}$]~Dyson equation solved using $\Sigma^{\rm (scr)}$,
density $\rho _{ep}$ from equation~(\ref{Str2}).
\end{indented}
\end{table}

The positron binding energy is highest in PsH. This is a consequence
of the small size of the hydrogen atom, and the small value of its
electron affinity, which makes for strong electron-positron correlation
effects. Beyond PsH the binding energy decreases along the halogen sequence,
mostly due to a stronger positron repulsion from the positively-charged
atomic cores in heavier systems.

Values of the normalization parameter $a$ in table~\ref{table2} give some
insight into the structure of these compounds. PsH has the smallest value of
$a$ and its structure therefore has a large component that describes Ps
bound to the neutral atom (``Ps cluster''), the small electron affinity of
H playing a role in this. PsF has the largest value of $a$ and its structure 
can best be described as a positron bound to F$^-$. Generally, all of the
compounds considered have large values of $a$. This indicates that a positron
bound to the negative ion is the dominant component of the structure.
This is a consequence of the stable noble-gas-like structure of the halogen
negative ions. In contrast, positron bound states with the weakly-bound
alkali negative ions have a distinct Ps-atom character (Mitroy \etal 2002).

\begin{figure}[ht]
\begin{center}
\includegraphics*[width=9.0cm]{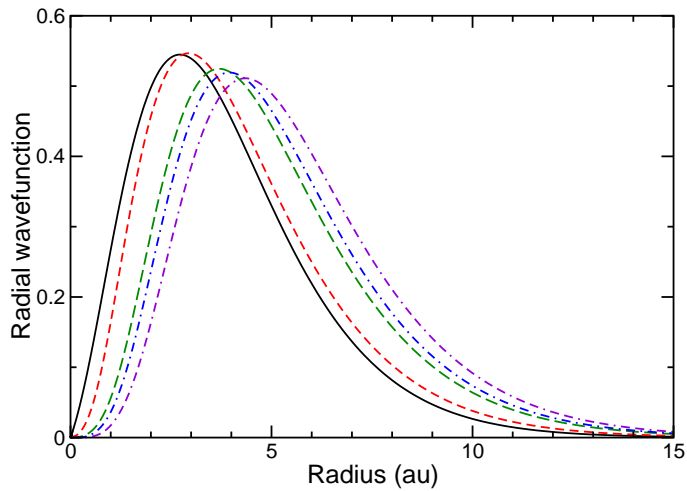}
\end{center}
\caption{Comparison of the radial Dyson orbitals obtained with
$\Sigma^{(2+\Gamma+3)}$: solid, PsH; dashed, PsF; long-dashed,
PsCl; dot-dashed, PsBr; dot-double-dashed, PsI. For the purpose of comparison,
all orbitals are normalized to unity.}
\label{fig:haldys}
\end{figure}

It is interesting to compare the positron wavefunctions obtained from
the Dyson equation for H$^-$ and the halogen anions. In figure~\ref{fig:haldys}
the wavefunctions obtained with the self-energy $\Sigma^{(2+\Gamma+3)}$ are
shown. 
The shape of the positron wavefunction is determined by a balance between the
Coulomb and correlation-potential attraction at large separations, and
the Coulomb repulsion from the nucleus at smaller radii.
The positron wavefunctions for PsH and PsF are quite similar. This feature
reflects the high positron binding energy to the H$^-$ and F$^-$ ions, and the
fact that the corresponding atoms have the smallest radii. As the
positron binding energy decreases, the positron wavefunction relaxes outwards.
This feature is seen as one moves along the halogen sequence. One can also
observe the increasing ``expulsion'' of the positron from the atomic core
region, caused by the Coulomb repulsion from the nucleus. 

\subsection{Comparison with other theoretical results}
The positron binding energies can be converted to Ps binding energies via the 
simple relation,
\begin{equation}\label{conv}
\mbox{BE}(\mbox{Ps}A)=\mbox{EA}(A)+\mbox{PA}(A^-)+E_{1s}(\mbox{Ps}) ,
\end{equation} 
where $\mbox{BE}(\mbox{Ps}A)$ is the binding energy of a Ps atom to a generic
atom denoted by $A$, $\mbox{EA}(A)$ is the electron affinity of the atom,
$\mbox{PA}(A^-)$ is the positron affinity of the negative ion and
$E_{1s}(\mbox{Ps})=-6.8028$~eV is the spin-averaged energy of Ps. For the
electron affinities needed, the values of 0.7542~eV for H,
3.4012~eV for F, 3.6127~eV for Cl, 3.3636~eV for Br and 3.0590~eV
for I, have been used (Andersen \etal 1999).

Annihilation rates $\Gamma_a$ in units of ns$^{-1}$ are obtained
by dividing the contact density $\rho_{ep}$ by the conversion factor
$10^9\,\mbox{s}/(\pi r_0^2c)=0.0198\,\mbox{ns}\times \mbox{au}$.
The final Ps binding energies and positron annihilation rates are
shown in table \ref{table3}, and compared with other calculations and
experiment. 

\begin{table}[ht]
\caption{\label{table3} Ps binding energies and positron annihilation rates 
for PsH and positronium halides compared with other
calculations and experiment.}
\begin{indented}
\item[]
\begin{tabular}{@{}lllll}
\br
Compound & \multicolumn{2}{c}{Present results} & \multicolumn{2}{c}{Other
results} \\
 & Ps BE (eV) & $\Gamma_a$ (ns$^{-1}$) & Ps BE (eV) & $\Gamma_a$ (ns$^{-1}$)\\
\mr
PsH & 1.118 & 2.544& 1.066126$^{\rm a}$&2.4361$^{\rm a}$  \\
&&&1.1$\pm$0.2$^{\rm i}$&\\
PsF & 2.718 & 2.482& 2.806$^{\rm b}$ & 2.019$^{\rm b}$   \\
&&&2.70$^{\rm c}$& 1.98$^{\rm c}$   \\
&&&2.24$^{\rm d}$&\\
&&&1.98$\pm$0.17$^{\rm g}$&\\
&&&2.838$^{\rm f}$&\\ 
&&&2.9$\pm$0.5$^{\rm j}$&\\   
PsCl & 2.245 & 1.984 & 2.350$^{\rm b}$ &1.504$^{\rm b}$\\
&&&1.91$\pm$ 0.16$^{\rm e}$&\\ 
&&&1.62$^{\rm d}$&\\ 
&&&2.0$\pm$0.5$^{\rm j}$&\\ 
PsBr & 1.873 & 1.913&2.061$^{\rm b}$ &1.371$^{\rm b}$\\
&&& 1.14$\pm$0.11$^{\rm g}$ &\\ 
&&&1.25$^{\rm h}$&\\ 
PsI & 1.393 & 1.809& 1.714$^{\rm b}$ & 1.254$^{\rm b}$\\
&&& 0.56$^{\rm h}$&\\ 
\br
\end{tabular}
\item[] Theory: $^{\rm a}$~Frolov \etal 1997, $^{\rm b}$~Saito 2005,
$^{\rm c}$~Miura and Saito 2003,
$^{\rm d}$~Saito 1995, $^{\rm e}$~Schrader \etal 1992a,
$^{\rm f}$~Bressanini \etal 1998, $^{\rm g}$~Schrader \etal 1993,
$^{\rm h}$~Saito \etal 1998.

\item[] Experiment: $^{\rm i}$~Schrader \etal 1992b,
$^{\rm j}$~Tao \etal 1969.

\end{indented}
\end{table}

For PsH very accurate variational calculations are available
(Frolov \etal 1997). Our many-body theory calculations are in good agreement
with these results, and both the Ps binding energy and positron annihilation
rate are accurate to within 5\%. Because of the small electron affinity of
hydrogen, PsH is more difficult for many-body theory to treat than larger,
more tightly bound systems with many valence electrons. The results for the
heavier systems should therefore be of similar or possibly even of greater
accuracy than the results for PsH.

For PsF a few theoretical calculations are available, the present Ps binding
energy and positron annihilation rate agreeing most closely 
with multi-reference configuration-interaction calculations
(Miura and Saito 2003, Saito 2005). The Ps binding energy is also close to a
diffusion Monte-Carlo calculation by Bressanini \etal (1998). 

The present Ps binding energies for PsCl, PsBr and PsI are greater than those
obtained using a second-order variational perturbation method
(Saito \etal 1995, 1998) and Monte-Carlo calculations (Schrader \etal 1992a,
1993). Our values are in better agreement with, although consistently smaller
than, multi-reference configuration-interaction calculations by Saito (2005).
The positron annihilation rates from the present calculation and that
of Saito (2005) are in reasonable agreement, although our values are
consistently higher.

\subsection{Comparison with experiment}
For PsH, a direct experimental measurement of the Ps binding energy
(Schrader \etal 1992b) is available. In this experiment, the reaction
$e^+ + \mbox{CH}_4 \rightarrow \mbox{CH}_3^+ + \mbox{PsH}$ was studied by
detecting the CH$_3^+$ ions. From the experimentally determined threshold
energy for CH$_3^+$ production and the various bond energies, a value of
$1.1\pm 0.2$ eV~was found for the PsH binding energy. This value is in
excellent agreement with theory, though much less accurate.

So far, there have been no direct experimental measurements of the Ps binding
energy for the halogens, however estimates of the Ps binding energy for
PsF and PsCl has been made (Tao \etal 1969), see table \ref{table3}.
The PsCl binding energy was estimated by studying positron annihilation in
Cl$_2$ and Ar-Cl$_2$ gas mixtures. The appearance of a shoulder in the
positron annihilation lifetime spectrum was attributed to the reaction,
$\mbox{Ps} + \mbox{Cl}_2 \rightarrow \mbox{PsCl} + \mbox{Cl}$. From a
knowledge of the energy at which this shoulder begins and the Cl$_2$
dissociation energy, a binding energy of about 2.0~eV was estimated for PsCl.
The estimate of the PsF binding energy was obtained from the observation that
when a hydrogen atom in benzene is replaced by fluorine, the fraction of
positrons annihilating with the longest lifetime, as ortho-positronium, was
reduced from 40 to 27\%. This reduction was assumed to be due to the reaction,
$\mbox{C}_6\mbox{H}_5\mbox{F} + \mbox{Ps} \rightarrow \mbox{PsF}
+ \mbox{C}_6\mbox{H}_5$. From a knowledge of the threshold energy and the
relevant dissociation energies, a binding energy of about 2.9~eV was estimated
for PsF. The present results support these early estimates.   


\section{Concluding remarks}\label{Concl1}

Traditionally, many-body theory has had more success in treating purely 
electronic systems (Chernysheva \etal 1988, Dzuba and Gribakin 1994), than
systems that contain a positron. In particular, earlier many-body theory
calculations relied on very simple (Amusia \etal 1976) or approximate
(Dzuba \etal 1995, 1996) treatments of the virtual Ps formation contribution
to the correlation potential. However it is now clear that many-body theory
is capable of giving positron and Ps binding energies and positron
annihilation rates that are accurate to within a few percent for
many-electron systems.  

The calculated binding energies and annihilation rates for positronium halides
should serve as a useful reference for other theoretical calculations and
future experiments. We have also performed an extended analysis of various
contributions to the correlation potential and contact density, especially of
the role of screening. This will be helpful for the problem of positron
scattering and annihilation on noble-gas atoms. In the future it will be
important to calculate explicitly the contribution of the RPA-type screening
diagrams and also to move to a fully relativistic framework, particularly for 
heavy atoms and ions.

\ack
The work of JL has been supported by DEL.

\section*{References}

\begin{harvard}

\item[] Amusia~M~Ya and Cherepkov~N~A 1975 Case Studies in Atomic Physics
{\bf 5} 47

\item[] Amusia M Ya, Cherepkov N A, Chernysheva L V and Shapiro S G 1976
{\it J. Phys. B: At. Mol. Phys.} {\bf 9} L531

\item[] Andersen~T, Haugen~H~K and Hotop~H 1999
{\it J. Phys. Chem. Ref. Data.} {\bf 28} 1526

\item[] Bell~J~S and Squires~E~J 1959 {\it Phys. Rev. Lett.} {\bf 3} 96

\item[] Bressanini~D, Mella~M and Morosi~G 1998 {\it J. Chem. Phys.}
{\bf 108} 4756

\item[] Cade~P~E and Farazdel~A 1977 {\it J. Chem. Phys.} {\bf 66} 2598

\item[] Chernysheva~L~V, Gribakin~G~F, Ivanov~V~K and Kuchiev~M~Yu 1988
{\it J. Phys. B: At. Mol. Opt. Phys.} {\bf 21} L419 

\item[] Danielson~J~R, Young~J~A and Surko~C~M 2009 {\it J. Phys. B:
At. Mol. Opt. Phys.} {\bf 42} 235203

\item[] Dzuba~V~A and Gribakin~G~F 1994 {\it Phys. Rev. A} {\bf 49} 2483

\item[] Dzuba~V~A, Flambaum~V~V, Gribakin~G~F and King~W~A 1995
{\it Phys. Rev. A} {\bf 52} 4541

\item[] Dzuba~V~A, Flambaum~V~V, Gribakin~G~F and King~W~A 1996
{\it J. Phys. B: At. Mol. Opt. Phys.} {\bf 29} 3151

\item[] Frolov~A~M and Smith~V~H~Jr 1997 {\it Phys. Rev. A} {\bf 55} 2662

\item[] Gilbert~S~J, Barnes~L~D, Sullivan~J~P and Surko~C~M 2002
{\it Phys. Rev. Lett.} {\bf 88} 043201

\item[] Gribakin~G~F, Gul'tsev~B~V, Ivanov~V~K and Kuchiev~M~Yu 1990
{\it J. Phys. B: At. Mol. Opt. Phys.} {\bf 23} 4505

\item[] Gribakin~G~F 2000 {\it Phys. Rev. A} {\bf 61} 022720

\item[] Gribakin~G~F 2001 {\it New Directions in
Antimatter Chemistry and Physics} ed C M Surko and F A Gianturco
(The Netherlands: Kluwer Academic Publishers) p~413

\item[] Gribakin~G~F and Ludlow~J 2002 {\it J. Phys. B: At. Mol. Opt. Phys.}
{\bf 35}, 339

\item[] Gribakin~G~F and Ludlow~J 2004 {\it Phys. Rev. A} {\bf 70} 032720

\item[] Gribakin~G~F, Young~J~A and Surko~C~M 2010 {\it Rev. Mod. Phys.}
submitted

\item[] Johnson W R, Sapirstein J and Blundell S A 1989 {\it J. Phys. B: At.
Mol. Opt. Phys.} {\bf 22} 2341

\item[] Migdal~A~B 1967 
{\it Theory of Finite Fermi-Systems and Applications to Atomic Nuclei}
( New York: Interscience)

\item[] Mitroy J, Bromley M W J and Ryzhikh G G 2001 {\it New Directions in
Antimatter Chemistry and Physics} ed C M Surko and F A Gianturco
(The Netherlands: Kluwer Academic Publishers) p~199

\item[] Mitroy J, Bromley M W J and Ryzhikh G G 2002 {\it J. Phys. B: At.
Mol. Opt. Phys.} {\bf 35} R81

\item[] Mitroy J and Bromley M W J 2006 {\it Phys. Rev. A} {\bf 73} 052712

\item[] Mitroy J and Bromley M W J 2007 {\it Phys. Rev. Lett.} {\bf 98}
063401

\item[] Mitroy J, Bromley M W J and Varga K 2007 {\it Phys. Rev. A} {\bf 75}
062505

\item[] Miura~N and Saito~S~L 2003 {\it Mol. Phys.} {\bf 101} 143

\item[] Ore~A 1951 {\it Phys. Rev.} {\bf 83} 665

\item[] Ryzhikh G G and Mitroy J 1997 {\it Phys. Rev. Lett.} {\bf 97} 4123

\item[] Saito~S~L 1995 {\it Chem. Phys. Lett.} {\bf 245} 54

\item[] Saito~S~L and Hidao~S 1998 {\it Chem. Phys. Lett.} {\bf 288} 277

\item[] Saito~S~L 2005 {\it J. Chem. Phys.} {\bf 122} 054302

\item[] Sapirstein~J and Johnson~W~R 1996 {\it J. Phys. B: At. Mol.
Opt. Phys.} {\bf 29} 5213

\item[] Schrader~D~M {\it Nucl. Instrum. Methods B} {\bf 143} 209

\item[] Schrader~D~M, Yoshida~T and Iguchi~K 1992a {\it Phys. Rev. Lett.}
{\bf 68} 3281

\item[] Schrader~D~M, Jacobsen~F~M, Frandsen~N~P and Mikkelsen~U 1992b 
{\it Phys. Rev. Lett.} {\bf 69} 57

\item[] Schrader~D~M, Yoshida~T and Iguchi~K 1993 {\it J. Chem. Phys.}
{\bf 98} 7185

\item[] Schrader~D~M and Moxom~J 2001 {\it New Directions in Antimatter
Chemistry and Physics} ed C M Surko and F A Gianturco
(The Netherlands: Kluwer Academic Publishers) p~263

\item[] Simons~L 1953 {\it Phys. Rev.} {\bf 90} 165 

\item[] Tao~S~J and Green~J~W 1969 {\it J. Phys. Chem.} {\bf 73} 882

\end{harvard} 
\end{document}